# Morphological Reconstruction Improves Microvessel Mapping in Super-Resolution Ultrasound

Scott Schoen Jr, Zhigen Zhao, Ashley Alva, Chengwu Huang, Shigao Chen, and Costas Arvanitis

*Abstract*—Generation of super-resolution (SR) ultrasound (US) images, created from the successive localization of individual microbubbles in the circulation, has enabled the visualization of microvascular structure and flow at a level of detail that was not possible previously. Despite rapid progress, tradeoffs between spatial and temporal resolution may challenge the translation of this promising technology to the clinic. To temper these trade-offs, we propose a method based on morphological image reconstruction. This method can extract from ultrafast contrast-enhanced ultrasound (CEUS) images hundreds of microbubble peaks per image (312-by-180 pixels) with intensity values varying by an order of magnitude. Specifically, it offers a fourfold increase in the number of peaks detected per frame, requires on the order of 100 ms for processing, and is robust to additive electronic noise (down to 3.6 dB CNR in CEUS images). By integrating this method to a SR framework we demonstrate a 6-fold improvement in spatial resolution, as compared to CEUS, in imaging chicken embryo microvessels. This method that is computationally efficient and, thus, scalable to large data sets, may augment the abilities of SR-US in imaging microvascular structure and function.

## I. INTRODUCTION

MICROBUBBLE ultrasound contrast agents (USCAs) exploit the way in which sound waves are reflected due to acoustic impedance differences between these microscale gas pockets and blood or tissue. As the acoustic impedance of gas is typically thousands of times lower than blood or tissue, gas microbubbles are ideal point scatterers for ultrasound, offering sufficient SNR to detect a single microbubble [1]. Moreover, their nonlinear response provides the means to isolate their echoes from the tissue. As such, USCAs are routinely used in the clinics for vascular imaging using B-mode (a pulse echo technique to localize linear microbubble echoes) or non-linear imaging, such as pulse inversion (a coherence-based phase inversion technique to localize nonlinear microbubble echoes) [2]–[4]. Despite the improved ability of contrast enhanced ultrasound to quantify tissue perfusion using USCAs, there is an inherent tradeoff between the resolution and penetration depth that, for clinically relevant frequencies, provides an effective lower bound to the imaging resolution (e.g. for a $f_0 = 4$ MHz pulse the limit is ~ $c/f_0 = 385$ μm).

By expanding concepts originally developed in super-localization optical microscopy [5], [6], it was shown that ultrasound in combination with microbubbles can alleviate tradeoffs between resolution and penetration depth [7]–[10]. Based on these methods, microbubbles that act as (moving) sono-activatable point sources are localized one by one with a precision far beyond the diffraction limit by identifying the centroid or peak intensity of each isolated peak region from each frame. Superimposing all the localized points to form density maps of USCA positions allows to resolve the microvascular "fingerprint" of organs with subwavelength resolution [10], [11]. Resolution eightfold below the diffraction limit has been consistently reported across several studies in healthy and diseased rodents using research or clinical ultrasound scanners [10]–[13] (see also Suppl. Table S-1). Such super-resolution ultrasound (SR-US) techniques have a theoretically-achievable resolution on the order of a few microns for clinical ultrasound frequencies [14].

However, the improved spatial resolution of SR-US comes at the cost of poor temporal resolution, defined as the total time required to acquire the data and generate the final image [15]. While high frame rates (up to several kilohertz) [16] have enabled reduced image acquisition times for SR-US from minutes [8], [10], [17] to a few seconds [7], [18]–[20], the processing time to generate the final image (i.e. peak extraction and final image formation) can take from several minutes to several hours to complete [15]. Moreover, due to i) the stringent selection and acceptance criteria (e.g. a single bubble detection threshold), ii) the proportionally fewer bubbles in microvessels, which are the vessels of particular interest in SR-US [21], and iii) the potentially lower SNR of the USCAs in these vessels (due to higher damping by the vessel wall), which might render them below detection thresholds [8], [9], [22], only a small fraction of the bubbles that fulfill all the selection criteria are used for imaging sub-diffraction vessels. Collectively these constraints limit our ability to overcome tradeoffs between temporal and spatial resolution with minimum penalty on image quality.

Optimizing each processing step in the SR framework (i.e., detection of microbubbles from the surrounding tissue,

Submitted 14 October 2020. This study was supported by the NIH Grants R00EB016971 (NIBIB) and R37CA239039 (NCI).
S. Schoen Jr (e-mail: scottschoenjr@gatech.edu) and Z. Zhao (e-mail: Zhigen.zhao@gatech.edu) are with the Woodruff School of Mechanical Engineering, Georgia Institute of Technology, Atlanta, GA USA 30332.
A. Alva (e-mail: ashley.alva@gatech.edu) is with the School of Electrical and Computer Engineering, Georgia Institute of Technology, Atlanta, GA USA 30332.
C. Huang (e-mail: huang.chengwu@mayo.edu) and S. Chen (e-mail: chen.shigao@mayo.edu) are with Department of Radiology, Mayo Clinic College of Medicine and Science,Mayo Clinic, Rochester, MN, USA 55905
C. Arvanitis is with the Woodruff School of Mechanical Engineering, Georgia Institute of Technology, and Coulter Department of Biomedical Engineering, Georgia Institute of Technology and Emory University, Atlanta, GA USA 30332 (e-mail: costas.arvanitis@gatech.edu).

isolation of individual microbubble signals, and localization of the microbubble at a precision beyond the diffraction-limit) could potentially address these tradeoffs without compromising image quality (i.e. how well a microvessel is resolved). For instance, increasing the number of bubbles identified per frame, could either reduce the number of frames required for the final SR-US image or lead to improved contrast-to-noise ratio (CNR; for a given number of frames) and accuracy, as more USCAs will be detected from a given number of frames. To date, up to tens of bubbles within a single frame have been detected (about 1 bubble per thousand pixels) using high SNR data sets [19], [20], while more typically tens of bubbles are found in each frame (See Suppl. Table S-1). Improving the number of bubbles detected in each frame, without introducing additional computational burden could have palpable effects on temporal resolution.

Here we propose to use grayscale morphological reconstruction (MR) to isolate and localize up to several hundred bubbles per frame with different signal intensities. We first analyze the ability of MR to super-localize points and characterize the effects of algorithm parameters and added noise on the generated SR images. With use of reference optical microscopy data, we estimate the accuracy of the localizations and the performance of the technique compared to the theoretical maximum performance. Finally, we measure the computational efficiency of the algorithm.

## II. METHODS

We used morphological image processing techniques to extract peak locations from the raw contrast enhanced image. We then used localization of many bubbles over time to generate super-resolution images, and evaluated the accuracy, sensitivity, and efficiency of the method.

### A. Data Acquisition

Data were collected as reported in [19] and as recounted briefly here. The CEUS images were taken from the chorioallantoic membrane (CAM) of chicken embryos. This model is attractive due to the long microbubble recirculation times, small bulk tissue motions, and the ability to directly compare the US images with high resolution optical microscopy images of the vasculature. A bolus injection of microbubbles (Lumason, Bracco Diagnostics Inc., Monroe Township, NJ) at $1.8 \times 10^9$ microbubbles per milliliter, and imaging acquisitions were performed at the microbubble concentration plateau after the injection.

CEUS images were obtained with a Verasonics Vantage 256 ultrasound system (Verasonics Inc., Kirkland, WA) with a 25 MHz linear array transducer (L35-16vX, Verasonics Inc.). Ultrafast plane wave imaging (15 angles, −7° to 7°) was performed at 500 compounded frames per second at a frame rate with 5 V transmit excitation. At each location, 5 successive acquisitions of 720 frames each (total acquisition length 3600 frames over 7.2 seconds). The IQ data were stored and post-processed with custom MATLAB (MathWorks, Natick, MA) scripts on a standard desktop computer (4 cores at 2.8 GHz, 16 GB RAM).

### B. Raw Data Filtering

To better distinguish the bubble signal from tissue scattering, raw CEUS images [Fig. 1(a)] an SVD filter was applied to each image stack. Singular values larger than 10% of the first (maximum) singular value (typically the first two to five values) were set to 0 [Fig. 1(b)]. Additionally, the smallest singular value was seen to contain largely noise and was also removed. Filtered frames were interpolated from the original pixel size (60 μm by 30 μm; diffraction limits in the axial and transverse directions) up to 12 times (5 μm by 2.5 μm). To ensure the intensity varied smoothly and improve isolation by the MR algorithm, the frames were then smoothed with a 2D Gaussian filter with size 30 μm.

### C. Morphological Reconstruction

To identify local intensity maxima whose absolute intensities may vary widely (i.e., that would be missed by simple thresholding), each filtered, smoothed, interpolated frame was multiplied by a factor $(1 - h)$ where $h \sim 0.05$, to obtain marker image, and the original map is used as the mask image [Fig. 1(c1)]. The optimal value of the morphological offset $h$ depends on the data; data with low background signal permits a smaller threshold and effectively higher sensitivity, which potentially may allow more bubble peaks to be found in each frame. Then, a grayscale morphological reconstruction was performed with the marker and mask images (via MATLAB's `imreconstruct`); this process may be considered as repeated dilations of the marker image until it fills the mask as in Fig. 1(c2). Finally, this reconstruction (the masked dilated image) was subtracted from the original smoothed map. The effect of this procedure is to segment the image into regions containing regional maxima (termed "$h$-domes" by Vincent [23]; see Appendix), which now have comparable amplitudes despite their disparate intensity values in the raw image [Fig. 1(c3)]. Finally, peaks in the reconstructed image with amplitudes within 10% of the peak intensity were retained. To compare with a simple thresholding method, the SVD filtered images were also binarized (such that image regions with less than 90% of the peak intensity were set to 0) and peak regions extraverted similarly.

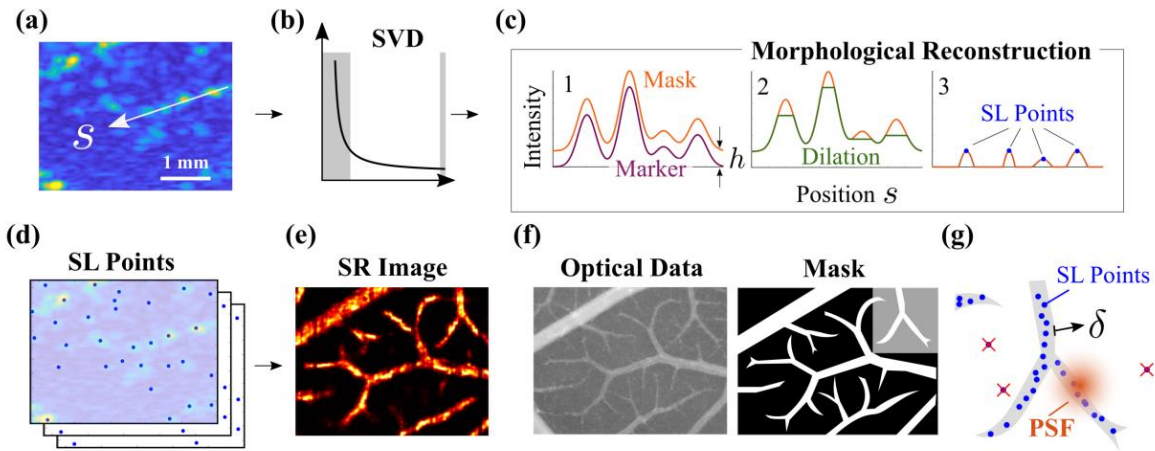

**Fig. 1.** Super-resolution ultrasound with morphological reconstruction and vascular characterization (**a**) Contrast-enhanced ultrasound (CEUS) were acquired with ultrafast plane wave imaging. (**b**) Singular value decomposition (SVD) filtering differentiates tissue from flowing contrast agents. (**c**) Morphological reconstruction with the intensity from the filtered frames scaled by $(1 − h)$ as the mask enables isolation of peak regions, and convolution with the PSF identifies super-localized (SL) points (blue dots). (**d**) Accumulation of SL point locations from all frames and (**e**) superimposed Gaussian profiles centered at these locations produces the super-resolution image. (**f**) Optical microscopy data of the same vasculature enables definition of a binary mask, on which (**f**) the SL points may be considered successful (if within the vessel) or spurious (if they are outside, red x's).

Finally, to super-localize the bubble within each segmented region, the orientation of each region was computed, and a local 2D convolution of the region was performed with a Gaussian approximation of the observed point spread function at the specific location (standard deviation of 30 µm in each direction) [24]. While a uniform PSF was used here, a position-dependent one could be used since the locations of the peak regions are known. The peak of each of these convolutions was taken as the super-localized bubble location [Fig. 1(c3)]. SR images were formed by summing uniform amplitude Gaussians (standard deviations of $\lambda/8$) at each super-localized bubble location [Fig. 1(d-e)].

The contrast-to-noise ratio (CNR) for the images at a given location $r$ was computed as

$$\text{CNR}(r) = \frac{I(r) - \mu_{\text{bg}}}{\sigma_{\text{bg}}}, \quad (1)$$

where $I$ is the image intensity, $\mu_{\text{bg}}$ and $\sigma_{\text{bg}}$ are the mean and standard deviation, respectively, of the intensity in a manually specified background region. To evaluate the performance of MR super-localizations in the presence of noise, we added various amounts of Gaussian noise, with amplitudes up to twice the mean intensity of the raw image.

### D. Evaluation of Localization Accuracy

To determine the accuracy and quality of the peaks found via MR, the final SR image (i.e., the intensity field due to the summed Gaussian distributions at each SR peak location) was registered with an optical microscopy image of the vasculature (via MATLAB's `imregister`). From the registered optical data, a binary mask was created as a reference standard [Fig. 1(f)], such that SR points in the acoustic image within the mask are considered true positives, i.e., they fall within the vasculature and may be considered microbubble localizations. As the registration is imperfect, a tolerance distance $\delta$ was defined such that if the SL points were less than $\delta$ from the vessel mask, they were considered successful localizations. Finally, knowledge of the mask area and size of the imaging wavelength were used to estimate the upper bound on the number of bubbles that could be localized (i.e., the area of the vessel mask divided by the size of a square wavelength). Finally, to confirm that peaks isolated by MR were due to flowing scatterers, we considered a simple nearest-neighbor pairing of localized peaks between frames to obtain estimation of the flow velocity; i.e., $\boldsymbol{v}_i = (\boldsymbol{r}_i^{n+1} - \boldsymbol{r}_i^n)/\Delta t$, where $\boldsymbol{r}_i^n$ is the position of peak $i$ in frame $n$, and $\Delta t$ = 2 ms was the time between the acquired frames.

## III. RESULTS

### A. Integration of Morphological Reconstruction into Super-Resolution Framework

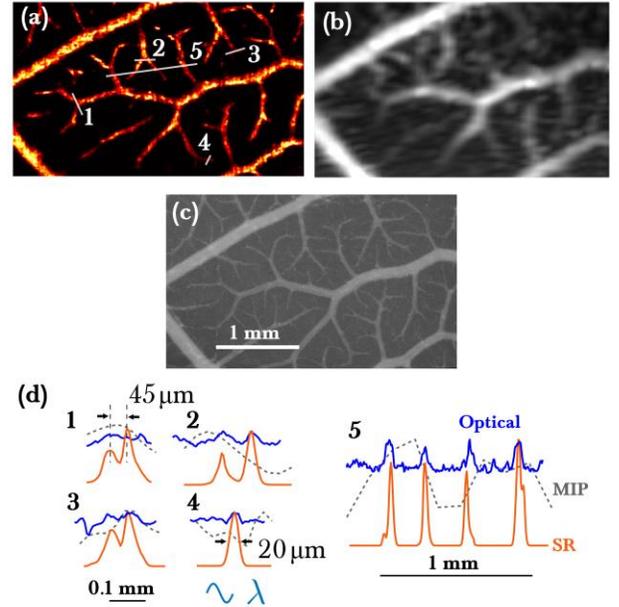

**Fig. 2.** Super-resolution with points recovered via morphological reconstruction. **(a)** Super-resolution image of vasculature with peaks found from MR ($h = 0.050$). **(b)** Maximum intensity projection of the CEUS image stack. **(c)** Registered optical microscopy of the corresponding vessel region. **(d)** Profiles at the labeled locations in (a) for the optical (orange) and super-resolved acoustic (orange) images, as well as from the maximum intensity projection of the CEUS images (gray). The wavelength (60 µm) is shown for reference.

To evaluate the integration of MR peak finding algorithm in the SR-US framework, we applied it to the ultrafast CEUS vascular imaging data sets of healthy CAM chicken embryos. Figure, 2(a) is the SR image based built from the peaks obtained via MR with $h = 0.050$. The image resolution is significantly better than that obtained from a maximum intensity projection of the CEUS stack [Fig. 2(b)] and is comparable to that of the optical image [Fig. 2(c)]. The intensity profiles shown in Fig. 2(d) for the indicated lines in Fig. 2(a) show that bifurcations with separations as small as 45 µm may be imaged (profile 1), and sub-vessel detail for vessels as small as 20 µm are identifiable (profile 4). These details, which are not visible in the CEUS image [gray lines in Fig. 1(d)], are resolved with high contrast in the SR image. Thus, the peaks identified with

MR may generate SR images with resolution significantly better than that of the raw CEUS images and comparable to that of optical microscopy.

### B. Sensitivity vs Accuracy of Super-Localized Peaks

After demonstrating that MR can be used for peak detection in SR-US imaging framework, we assessed its robustness and accuracy. As the co-registered optical data is available, we consider these data as a ground truth, and compared the super-localized peaks obtained from the MR process with a binary mask created from the vascular map. The location of each super-localized point was compared with the binary mask. Peaks that fell within the vessel region (i.e., at pixels where the mask had value 1) were labeled as within the vessel (i.e. real peaks), and outside the vessel otherwise (i.e. faulty peaks). To account for image registration errors different tolerances (i.e. vessel true location) were considered.

Figure 3 demonstrates that without MR, $20.5 \pm 3.4$ peaks were detected in each frame, with 89.6% located within the vessel mask. For a tolerance of 20 µm, 95.7% were within the vessel, and for a 50 µm, 98.6% were labeled as within the vessel. The peaks found per frame via MR were approximately two to three times as many: for an offset $h = 0.075$, $38.8 \pm 4.6$, for $h = 0.050$, $49.1 \pm 4.8$, and for $h = 0.025$, $66.2 \pm 5.3$. The increased sensitivity ($h = 0.025$) had slightly lower accuracy than the non-MR case; for instance, the $h = 0.025$ had a lower bound ($\delta = 0$ µm) of 69.7% localized within the vessel. However, given the imperfect registration (see Suppl. Fig. S-1), for even a small tolerance of 20 µm labeled 86% of these localizations is within the vessel.

Finally, since the area of the vasculature is known, an approximate upper bound on the number of separable point sources per frame that might be identified by the diffraction-limited system can be established. If each source is imaged as a brightness peak with an area of 1 square wavelength, then the total area of the vasculature in Fig. 2 divided by the PSF area yields a theoretical maximum of 667 bubbles. Given that MR identifies up to 62 peaks per frame, approximately 10% of this upper bound was achieved.

The peak regions had a mean size of approximately 0.2 square wavelengths consistent with the assumption that the identified peaks are due to subwavelength scatterers, as the area of the peak region is governed by the spatial distribution of the peak, rather than by its amplitude or depth within in the raw image (see Supplementary Figs. S-2 and S-3). Finally, velocity estimations from the pairing of peaks between frames yield flow rates and directions consistent with more sophisticated algorithms [20], suggesting the localizations are indeed flowing contrast agents (see Suppl. Fig. S-4).

Collectively our findings indicates the potential abilities of the MR algorithm for accurate (number of bubbles within the vessel) and precise (fraction of bubbles within the vessel) microbubble peak detection allowing to assemble SR images with resolution comparable to optical imaging.

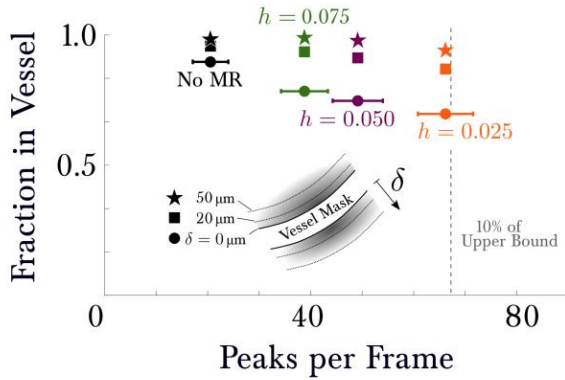

**Fig. 3** Accuracy vs Sensitivity for MR peak finding. Without MR (black markers) approximately 20 peaks were localized in the region shown in Fig. 2. For a small offset (higher sensitivity, orange markers) many more peaks were detected in each frame, though the fraction of these peaks that were within some tolerance $\delta$ of the vessel mask decreased slightly. The theoretical upper bound on the number of peaks that could be detected was approximately 10 times the value near the dashed gray line.

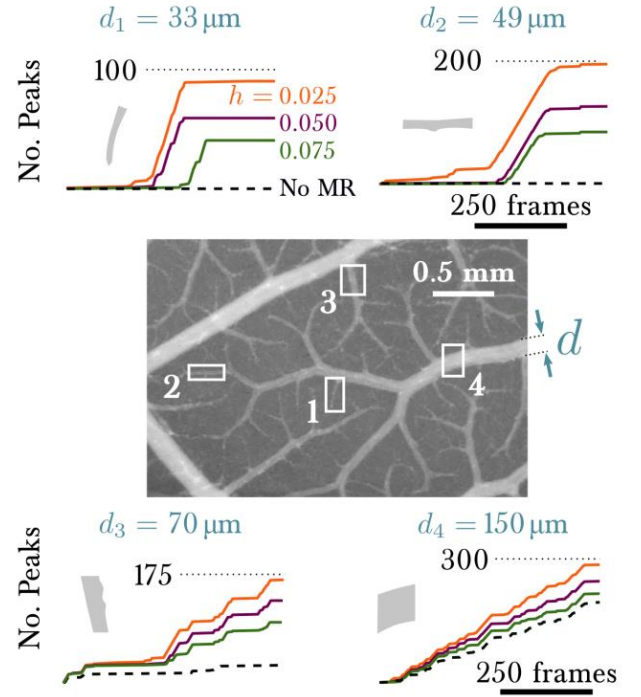

**Fig. 4** Effect of morphological offset on sensitivity. For each region in the optical microscopy image of the vasculature, the number of peaks detected in the first 720 frames (1.2 s) are shown in each subplot, as well as an outline of the vessel shape in gray. Results are shown with offsets of $h = 0.025$ (orange), 0.050 (purple), and 0.075 (green) as well as without morphological reconstruction (dashed black lines).

### C. MR Improves Localization in Small Vessels

To assess the abilities of MR to identify USCA peaks within small vessels, which are also the vessels of interest in SR-US and have low incidence rate, we evaluated the number of peaks found in different vessels as a function of time (Fig. 4). In the smallest vessels ($d_1 = 33$ µm and $d_2 = 49$ µm), no peaks were detected via thresholding (i.e., without MR and intensity threshold of 0.9). However, tens to hundreds of peaks were identified via MR, with more peaks isolated for lower offsets (higher sensitivity). For larger vessels, thresholding identified more bubbles, though still the number of localizations was significantly fewer than the number identified with MR. Together these data emphasize the potential of MR to identify USCA peaks within small vessels.

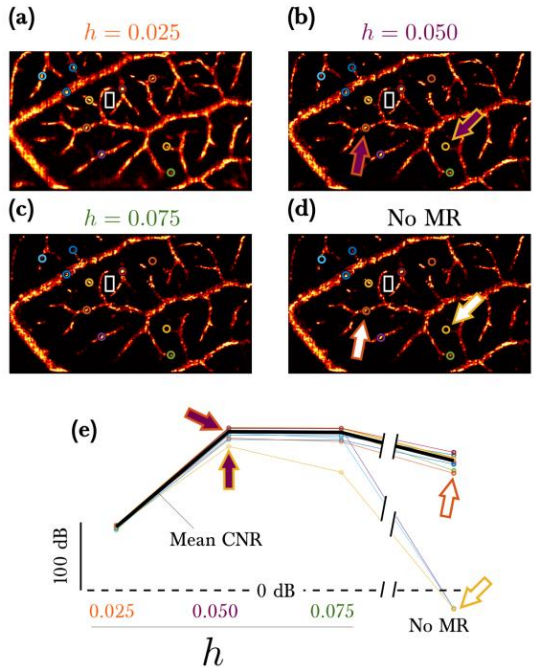

**Fig. 5** Effect of morphological offset on sensitivity. For each region in the optical microscopy image of the vasculature, the number of peaks detected in the first 720 frames (1.2 s) are shown in each subplot, as well as an outline of the vessel shape. Results are shown with offsets of $h = 0.025$ (orange), 0.050 (purple), and 0.075 (green) as well as without morphological reconstruction (dashed black lines).

### D. Higher Sensitivity Maintains Contrast

To ensure that the additional peaks detected were not spurious localizations we evaluated the contrast in the resulting SR images (Fig 5). While for the smallest offset ($h = 0.025$), the CNR was lower than the contrast in the non-MR images (i.e., the image formed with only thresholded peaks), likely due to some spurious localizations contributing to background noise, for larger offsets ($h = 0.050$), the CNR was quite high and exceeded that of the non-MR images. This is because the MR was able to identify much more peaks in the smaller vessels, resulting in stronger signal as compared to the same positions in the non-MR images [Fig. 5 (e)].

To demonstrate the robustness of the resultant images to measurement noise, we evaluated the CNR for images formed with $h = 0.05$, over the same vessel and background locations shown in Fig. 5. Figure 6 demonstrates that while the addition of noise (resulting in a mean CNR in the maximum intensity projection of the raw image stack if $3.6 \pm 10.0$ dB over the same locations) to the CEUS frames decreases the CNR of the output SR image, the contrast between vessels is consistent across all locations and remains positive [Fig. 6(b)]. However, without MR, some smaller vessels have poor contrast [as in Fig. 5(e)], which give vanishing CNRs and subsequently wide variability in the image contrast between different size vessels.

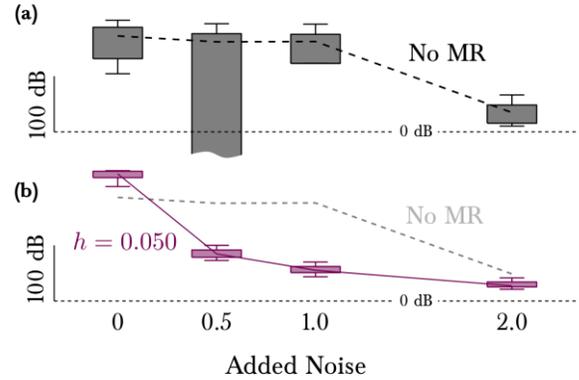

**Fig. 6** CNRs averaged over the locations shown in Fig. 5 computed for images formed with the indicated amount of added noise. **(a)** CNRs for images formed with localizations found without MR. **(b)** CNRs for images formed with localizations with MR and $h = 0.05$. Lines represent median values, whiskers the data range, and boxes the middle 50th percentile.

### E. MR is Computationally Efficient

Finally, we measured the computation time required per frame of the MR peak finding algorithm as a function of the interpolation of a 1 mm by 1 mm region from the CAM dataset shown in Fig. 7 (times do not include SVD filtering, which required 6.6 s for each 720 frame stack). Smaller offset cases required slightly longer processing times (e.g., $18.9 \pm 4.0$ ms vs $16.6 \pm 2.8$ ms per frame at 4x interpolation for $h = 0.025$ and $h = 0.075$, respectively) due to the larger number of points found (Fig. 3) in each frame, which increases number of convolutions. Peak finding without MR required slightly less time ($10.4 \pm 4.0$ ms per frame), but identified only 5.2 peaks per 1000 pixels over the same region, compared to 20 in the MR case. Thus, MR enables a roughly two-fold improvement in the temporal resolution (defined as the total acquisition and processing time to generate the SR image) for these data.

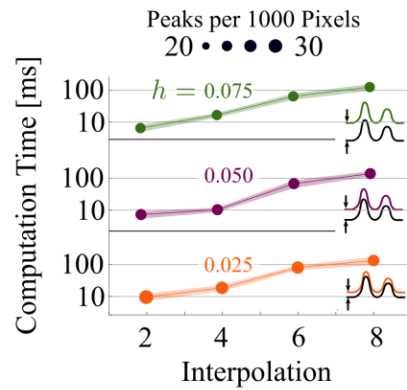

**Fig. 7** Effect of morphological offset on sensitivity. For a 1 mm by 1 mm region of the vasculature, the number of peaks detected in the first 720 frames (1.2 s) and computation times for each interpolation are indicated by the marker size and $y$-value, respectively. Results are shown with offsets of $h = 0.025$ (orange), 0.050 (purple), and 0.075 (green).

## IV. Discussion

Imaging the microvasculature noninvasively and beyond the capabilities of conventional ultrasound systems has driven the development of a range of super-resolution techniques. This paper describes a computationally efficient method for microbubble isolation and localization based on morphological reconstruction and demonstrates that it can be readily integrated to SR-US framework (Fig. 1). The super-localized points formed SR images that resolved the vasculature of a chicken embryo with resolution comparable to optical microscopy, and substantially better than a maximum intensity projection (Fig. 2). According to the optical microscopy images of the vasculature, approximately 90% of the localizations corresponded to positions within vessels (Fig. 3 and Suppl. Fig. S-1). Moreover, the number of peaks approached 10% of the maximum possible number of points given the size of the PSF and total vascular area. The CNRs of the SR images obtained with the super-localized peaks were higher at small vessels as compared to images formed with the thresholded peaks (Fig. 5). Finally, the peak finding routine was quite efficient, requiring, e.g., less than 10 ms per frame of a 1 mm by 1 mm region at 2x interpolation (65-by-130 pixels), in which up to 30 peaks were isolated (Fig. 7).

Existing methods for microbubble localization have reported identification of a few to several tens of microbubbles [7], [8], [17], [25], [19], [20], [26], [27] (see Suppl. Table S-1). The MR-based localization procedure presented here identified up to $66 \pm 5$ total intensity peaks per frame for the dataset in the region shown in Fig. 2 (75 by 65 pixels native resolution), a three-fold increase compared to the thresholded case without MR, while being nearly as accurate (94% within vasculature as determined by optical comparison vs 98% without MR). The improvement over the non-MR case was most pronounced in smaller vessels (less than 50 μm), for which no localizations were found via thresholding, but for which several hundred were with MR (Fig. 3). While a lower threshold (higher sensitivity) may detect more peaks in such vessels, the likelihood of false positives is also increased, and would require more careful processing to remove false localizations.

A significant benefit of the MR algorithm is that it is agnostic to imaging modality and thus amenable to analysis of both B-mode and PI images. Moreover, the computational efficiency of the algorithm (order of 100 ms per frame for the regions analyzed here) allows relatively efficient determination of multiple points. As the computation for each frame is independent, the algorithm is also fully parallelizable. Additionally, morphological operations are naturally extensible to three dimensions and could thus be used for volumetric imaging data [23], which would obviate the problem of our-of-plane localization errors. Information about the peak region geometries may also be of interest for deep learning-based methods, e.g., to discriminate overlapping bubble signal [28]. More research in this direction is warranted.

The MR method for peak detection described here has a few limitations. First is the choice of parameters. For MR-based peak finding, the offset $h$ must be empirically determined for the data set, depending on the SNR of the data. A range of 0.025 to 0.075 was seen to give good results here, though larger values may be required for data with higher background signal. Additionally, given the relatively high USCA concentration used in the current experiments, it cannot be stated with certainty that the super-localized peaks represent signal from a single bubble. However, the area of the peak regions found in the reconstructed images suggest that the scatterers are appreciably smaller than a wavelength (see Suppl. Figs. S-2 and S-3) and that they can be used to estimate vascular flow (Suppl. Fig. S-4), and their localizations were predominantly within the vasculature (Fig. 3 and Suppl. Fig. S-1). Moreover, the ability of the presented method to support the localization of a densely spaced group of contrast agents may complicate tracking and velocity calculations. Future work with varying concentrations or simultaneous optical imaging of the bubbles may address this concern, potentially combined with automated selection of algorithm parameters and qualification of peak detection based on the statistical characteristics of the peak.

## V. Conclusion

We have presented how the morphological reconstruction algorithm can be integrated to the SR-US framework. Peak extraction via MR enables two to three-fold increase in the number of peaks detected per frame compared with a thresholding technique with comparable accuracy as determined by comparison to optical ground truth. Further, comparison with the vascular density of the model suggest that nearly 10% of the theoretical maximum number of localizations was achieved. The greater sensitivity of the method enables improved detection of peaks in small vessels, while maintaining good image contrast. The method requires on the order of 100 ms per frame for processing. Together, the proposed framework could augment our ability to perform SR-US and may facilitate the development of clinically effective SR-US.

## Appendix

After [23], consider discrete distributions $I(\boldsymbol{r})$ and $J(\boldsymbol{r})$ defined on a rectangular, connected domain $D$ (i.e., the 2D intensity profile represented by a grayscale image). Now define as $T_k(I)$ the set of points in $I$ whose intensity is larger than some threshold $k$:

$$T_k(I) = \{\boldsymbol{r} \in D | I(\boldsymbol{r}) \geq k\}. \quad (2)$$

The reconstruction of the $I$ (called the mask image) from $J$ (called the marker image) is denoted $\rho_I(J)$ and is defined

$$\rho_I(J)(\boldsymbol{r}) = \max\{\boldsymbol{r} \mid \boldsymbol{r} \in \rho_{T_k(I)}(T_k(J))\}. \quad (3)$$

An equivalent but perhaps more intuitive definition employs the dilation operation $\delta$. Given a structuring element $S(\boldsymbol{r}')$ where $\boldsymbol{r}' \in D' \supseteq D$, the dilation is defined

$$\delta(I) = I \oplus S \equiv \max_{\boldsymbol{r}' \in D}[I(\boldsymbol{r}) + S(\boldsymbol{r} - \boldsymbol{r}')]. \quad (4)$$

That is, every value of $I$ is replaced with the maximum value of $I + S$ within a neighborhood defined by local support of $S$. Typically, the template $S$ is defined as 0 for the 3-by-3 neighborhood, such that each pixel is replaced with the maximum value of any adjacent pixel. The dilation of $J$ under $I$, written $\delta_I(J)$ is then simply

$$\delta_I(J) = \delta(J) \wedge I, \quad (5)$$

where $\wedge$ indicates the point-by-point minimum is taken. This operation dilates $J$, but limits the intensity by the maximum value of $I$ at each position. Successive iterations of this dilation operation are denoted as $\delta_I^{(n)}(J) = \delta_I(\cdots \delta_I(J) \cdots)$. Then, the morphological reconstruction of $I$ from $J$ may be expressed

$$\rho_I(J)(\boldsymbol{r}) = \bigvee_n \delta_I^{(n)}(J), \quad (6)$$

where $\vee$ denotes the point-by-point maximum. In Eq. (6), the operation is repeated until the output stops changing. To extract the peak regions, the grayscale CEUS frame is used as $I$, the shifted distribution $I - h$ is used as $J$. The image containing peak regions $P$ (and that is 0 elsewhere) is

$$P = I - \rho_I(J). \quad (7)$$

See also Supplementary material and Suppl. Fig. S-5.

ACKNOWLEDGMENT

Drs. Levent Degertekin and Heechul Yoon provided helpful comments on our manuscript.